\documentclass[aps,pre,reprint,groupedaddress,showpacs]{revtex4-1}
\usepackage{amsmath,amsfonts,amssymb}
\usepackage{graphicx}
\usepackage{natbib}
\usepackage{hyperref}

\begin{document}

\title{Tracer dispersion in the turbulent convective layer.}

\author{Alex Skvortsov}
\email[]{alex.skvortsov@dsto.defence.gov.au}
\affiliation{Defence Science and Technology Organisation,  506
Lorimer Street, Fishermans Bend, Vic 3207, Australia}

\author{Milan Jamriska}
%\email[]{milan.jamriska@dsto.defence.gov.au}
\affiliation{Defence Science and Technology Organisation,  506
Lorimer Street, Fishermans Bend, Vic 3207, Australia}

\author{Timothy C. DuBois}
%\email[]{timothy.dubois@dsto.defence.gov.au}
\affiliation{Defence Science and Technology Organisation,  506
Lorimer Street, Fishermans Bend, Vic 3207, Australia}

\date{\today}

\begin{abstract}

Experimental results for passive tracer dispersion in the turbulent surface layer under convective conditions are presented. In this case, the dispersion of tracer particles is determined by the interplay of two mechanisms: buoyancy and advection. In the atmospheric surface layer under stable stratification the buoyancy mechanism dominates when the distance from the ground is greater than the Monin-Obukhov length, resulting in a different exponent in the scaling law of relative separation of lagrangian particles (deviation from the celebrated Richardson's law). This conclusion is supported by our extensive atmospheric observations. Exit-time statistics are derived from our experimental dataset, which demonstrates a significant difference between tracer dispersion in the convective and neutrally stratified surface layers.

\end{abstract}

%Boundary layer turbulence, Statistical theories and models, Fractal and multifractal systems, time series analysis, Boundary layer structure and processes, Particles and aerosols

\maketitle

\section{Introduction}

Understanding transport properties of turbulent convective flow (e.g.\ transport of particles, chemical species, temperature, \emph{etc.}) is of significant importance for a number of fields of science and technology covering many physical applications including: environmental pollutant dispersion, heat transfer, combustion, reacting flows, meteorology and oceanology. Due to the buoyancy effect and the associated anisotropy, convective turbulence produces much more complex phenomenology than the classical Kolmogorov model of isotropic turbulence and this imposes new challenges for its analytical treatment \cite{Boffetta_12,Calzavarini_02,Toschi_00}. In spite of this complexity there has been remarkable progress in understanding this phenomena based on the application of modern methods of theoretical physics (see \cite{Frisch_96, Lohse_10, Celani_04} for comprehensive reviews). Examples include the derivation of Kolmogorov and Bolgiano spectra in thermal convection, scaling of statistical moments of concentration, multi-fractal structure of tracer statistics and some others (see \cite{Boffetta_08, Biferale_01, Calzavarini_02, Bistagnino_07, Bukai_11} and references therein).

An important step for further refinement of the developed theoretical framework is experimental validation of some of these theoretical predictions, which was motivation for the reported study. In this paper we present our experimental results on the dispersion of passive tracers in the turbulent convective layer.  The wall-bounded turbulent flow  under the effect of changing stratification provides flexible settings to study the relative contribution of the buoyant and kinetic energy fluxes on scalar transport since this contribution can be easily controlled by varying the distance to the underlying surface (and passing the threshold of the Monin-Obukhov length) \cite{Calzavarini_02, Bistagnino_07, Boffetta_08}. The results presented here are an extension of our previous study \cite{Skvortsov_10} conducted for the case of the neutrally stratified turbulent surface layer.

Various mechanisms of turbulent dispersion can be characterized by the mean inter-particle distance \cite{Frisch_96, Falkovich_01, Celani_04, Lohse_10}:
\begin{equation}
 \label{eqn:E:01}
\langle{R^{2}(t)}\rangle \sim  \lambda t^{n},
\end{equation}
where $n$ is the power exponent of the particle separation law and is specific for a particular mechanism of turbulent dispersion; $\lambda$ is a scale-independent dimensional parameter. A particular value of parameter $n$ can be associated with an energy injection mechanism of turbulence and can be used for universal characterization of associated dispersive properties. For instance, for tracer dispersion by Kolmogorov (locally isotropic) turbulence, $n = 3$ (Richardson's law) and for Bolgiano-Obukhov (buoyancy dominated) turbulence, $n = 5$ \cite{Bistagnino_07, Boffetta_08, Boffetta_12}. In the surface layer turbulence the dominant dispersion mechanism is associated with the strong gradient of dissipation energy leading to `ballistic' separation, $n = 2$ \cite{Skvortsov_10}.

Based on the above comments we can draw an important conclusion regarding the time evolution of parameter $n$ in the turbulent surface layer. More specifically, we infer that the long-time trends of this evolution  should exhibit a striking difference in the convective and neutrally stratified surface layer. Indeed, initially in both cases $n = 3$  (this is valid for the short-times which correspond to the short spatial scales, so turbulence can be considered as isotropic). Then in the neutrally stratified turbulent surface layer $n$ \emph{decreases}, approaching the global ballistic asymptote $n = 2$ \cite{Skvortsov_10}. Contrarily, in the convective layer $n$ \emph{increases} reaching its `convective' asymptote $n = 5$. Since a relative increase (or decrease) of parameter $n$ is quite significant, this difference should eventually manifest itself in experimental observations and the statistics of tracer particles. Observing this phenomenon was the rationale behind and one of the main goals of the current study.

The exponent $n$ can be estimated from time series data of particle concentrations, as there is a well-known connection between the power exponent of the particle separation law  and passive tracer statistics \cite{Frisch_96, Falkovich_01}:
\begin{eqnarray}
\label{e:assimp_eq}
 S_2(R) = 2[F_2(0) - F_2(R)] \propto R^{2/n},
\end{eqnarray}
where
\begin{eqnarray}
\label{e:assimp_eq1} F_2(R) =  \langle C (\textbf{x} + \textbf{R}, t), C (\textbf{x}, t)\rangle
\end{eqnarray}
is the pair-correlation function of our concentration levels, $C$. These expressions are used to estimate the value of $n$ in our model.

\section{Experimental Procedure}

The experimental procedure was fully described in our previous publications \cite{Skvortsov_10, Jamriska_12}. Consequentially, here we briefly present details that are specifically relevant for the reported set of new measurements.

Measurements of particle concentration were performed at a given location within the Taylor hypothesis of frozen turbulence $R = Ut$, where the wind speed $U = const$. From (\ref{eqn:E:01}), it follows that:
\begin{equation}
\label{eqn:E:04} S_2(R) = 2[F_2(0) - F_2(R)] \propto R^{2/n} \propto
(Ut)^{2/n},
\end{equation}
so single point time-measurements of concentration also show power-law behavior with the same characteristic exponents as the spatial measurements. From these observations we estimated the value of parameter $n$ in (\ref{eqn:E:04}).

A continuous (and relatively stable) influx of particles into the turbulent flow was supplied by anthropogenic activity in the surrounding urban areas (about $20$ km in size). The observation tower was at a height of $H = 12$ m and the tracer concentration was measured by means of scattered light intensity off sampled particles. Continuous observation was undertaken over a period of 21 days with an air sample reading taken every 5 seconds ($30$ liters per sample). The instrumentation provided total particle count as well as particle size distribution in the range of $1 - 10 \; \rm{\mu m}$. Local meteorological observations (three conventional meteorological stations in the vicinity) were available to draw conclusions about local meteorological conditions of the surface layer (stability, profiles of wind speed and temperature). Wind speed and temperature were also measured directly at the sampling point ($H = 12$ m). The Monin-Obukhov length $L_{mo}$ was estimated from these inputs \cite{Venkatram_04, Wyngaard_10}.

As usual in meteorological studies, we anticipated that the night time observations would correspond to the boundary layer under stable or neutral conditions. Contrarily, for convective turbulence conditions to be extant, an intensive heat flux from the underlying surface has to occur and this, of course, should include the significant effect of solar radiation. Therefore, for study of particle dispersion in the convective regime of the turbulent surface layer we focused our observations on hot, sunny days during the morning hours (i.e.\ the measured air temperature was always above $20^{\circ}$C). The data sets with negative values of $L_{mo}$ were selected with anticipation that additional constraints $|L_{mo}| < H$ and low wind velocity ensuring convective dominated mechanism of particles dispersion at sampling location.

\section{Results}

\begin{figure}[h]

\includegraphics[width=\columnwidth]{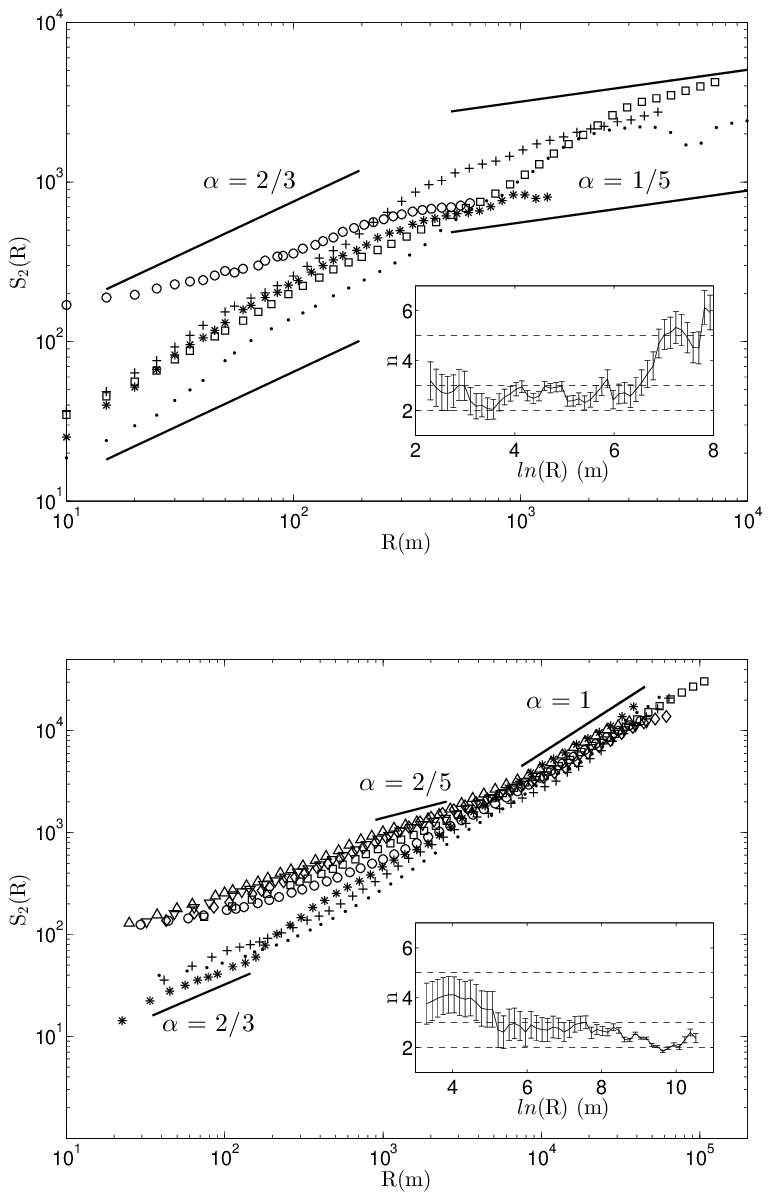}
%\subfloat{
%    \includegraphics[width=\columnwidth]{S2vRconv}
%}
%
%\subfloat{
%    \includegraphics[width=\columnwidth]{nightRcv}
%}

\caption{\label{fig:01}Structure functions of the tracer concentration $S_2(R) \propto R^{\alpha} \propto (U_{H}t)^{\alpha}$ on log-log scales: top plot - convective conditions; bottom plot - neutral conditions (as published in \cite{Skvortsov_10} for reference). Insets show convergence of the scaling exponent $n$ of the mean inter particle displacement converges over time ($\langle{R^{2}(t)}\rangle \propto t^{n}$). Each data set in the convective conditions plot corresponds to a maximum of four hours of observations, with wind velocities Monin-Obukhov lengths outlined in Table \ref{table:01}: $+$, Sample $1$; $\circ$, Sample $2$; $\ast$, Sample $3$; $\cdot$, Sample $4$ and $\square$, Sample $5$. $\alpha = 2/3$ is the Richardson regime, $\alpha \approx 1/5$ is the convective regime, $\alpha \approx 2/5$ is the velocity shear mechanism and $\alpha = 1$ is the ballistic regime. Global convergence to the convective regime ($\alpha = 1/5$) is clearly visible.}

\end{figure}

Undertaking correlation analysis of the time series of the concentration $C(t)$ of single point measurements, we found that the minimum time span that provides reliable statistics of the process should be more than two hours. As the first step, we plotted $S_2(R)$ as a function of $R = Ut$ on a log-log scale and evaluated the scaling exponent $n$ from the expression  $S_2(R) \propto (Ut)^{2/n}$ predicted by (\ref{e:assimp_eq1}) and (\ref{eqn:E:04}). Some examples of these plots are depicted in Fig \ref{fig:01}.

In this paper we report on the analysis of the data that corresponds to the convective conditions of the turbulent surface layer. Similar results observed for the neutrally stratified surface layer \cite{Skvortsov_10} are presented for comparison.

The values of mean wind speed $U_H$ and Monin-Obukhov length $L_{mo}$ relevant to the data series depicted in Fig.\ \ref{fig:01} are listed in Table \ref{table:01}, where $U_H = U$ at elevation from the ground, $H$ ($12$ m). The value of Monin-Obukhov length $L_{mo}$ was estimated from local meteorological station data for each observation period \cite{Venkatram_04, Wyngaard_10}. As mentioned above, for the hight $H > |L_{mo}|$ the turbulence in the surface layer is dominated by the convective flux, so we expect that the asymptotic value $n=5$ should always emerge. Contrarily, for the measurements in the non-convective scenario we anticipated to recover the global ballistic regime ($n = 2$) that is valid for tracer dispersion in the  neutrally stratified turbulent surface layer. The classical Richardson regime with $n = 3$ should be only a short-time asymptote in both cases. Particular scenarios approaching `global' asymptotes ($n = 2$ or $n = 5$) may be rather convoluted due to the shear dispersion mechanism, which can provide a profound contribution during intermediate-times \cite{Skvortsov_10}. However, at the long-time limit; parameter $n$ will eventually tend to either asymptote if the aforementioned stability conditions return to the turbulence surface layer.

\begin{table}[h]
     \caption{\label{table:01} Mean values for wind speeds $U_H$ and Monin-Obukhov length $L_{mo}$ as used in Fig. \ref{fig:01}}
\begin{ruledtabular}
\begin{tabular}{ccrcc}
& Sample & $U_H$ (m/s) & $L_{mo}$ (m) & \\
\hline
& 1. & $2.19\pm 0.39$ & $-9.36\pm 3.71$ & \\
& 2. & $1.20\pm 0.11$ & $-5.55\pm 1.32$ & \\
& 3. & $1.64\pm 0.23$ & $-7.43\pm 1.70$ & \\
& 4. & $4.00\pm 0.12$ & $-9.71\pm 1.87$ & \\
& 5. & $4.39\pm 0.32$ & $-8.80\pm 3.34$ & \\
\end{tabular}
\end{ruledtabular}
\end{table}

The existence of two significantly distinct asymptotes for particle separation clearly emerges from our experimental data (see Fig.\ \ref{fig:01}). More specifically, we observe that at the short-time intervals, $t \rightarrow 0$, the parameter $n \approx 3$ (i.e.\ the classical Richardson regime valid for isotropic turbulence), then it always reaches two significantly different limiting values (insets in Fig.\ \ref{fig:01}): $n \approx 5$ (dispersion in convective turbulence) or $n \approx 2$ (dispersion in neutrally stratified turbulence). The intermediate scenarios of approaching these asymptotes may exhibit non-monotonic behavior, but eventually either the convective or ballistic regime will dominate.

In order to further reveal the difference between particle transport in convective and neutrally stratified turbulent surface layers we employ the algorithm of exit-time statistics for the concentration time series. These are statistics of time intervals where a measured value of concentration exits through a set of thresholds, $\delta C$ \cite{Biferale_01, Celani_04}. By scanning the time series for a given threshold, one can recover a set of times $\tau_i(\delta C)$, where the measured concentration reaches this threshold. This set can then be used to calculate the Inverse Structure Function (for details see \cite{Celani_04}):
\begin{equation}
\label{eqn:E:06} \Sigma_q (\delta C) \equiv  \langle \tau^q(\delta
C) \rangle.
\end{equation}
where $q$ is a statistical moment.

A comprehensive analysis of the properties of the Inverse Structure Function can be fulfilled by applying the well-known multifractal approach \cite{Biferale_01, Boffetta_08, Schmitt_05}. This results in the following scaling:

\begin{equation}
\label{eqn:E:063} \Sigma_q (\delta C) \propto (\delta
C)^{\chi(q)},~~ \chi(q) = \min_h [{({q + 3 - D(h)})}/{h}],
\end{equation}
where $h$ is the index of singularity from the range $[h_{min}, h_{max}]$ such that $t \sim (\delta C)^{h}$ and $D(h)$ is the fractal dimension of the set with a singularity $h$. It is  worth noting that particular values $h=1/5$, $h=1/3$ and $h=1/2$ correspond to convective dispersion, the Richardson law and ballistic scaling respectively, as discussed above.

The function $(q + 3 - D(h))/h$ is assumed  to reach its minimum at the upper boundary of the singularities' range \cite{Schmitt_05},  so we can set $h=h_{max}$ and write
\begin{eqnarray}
\label{eqn:E:064}
 \chi (q) = (q + 3 - D(h_{max}))/h_{max}.
\end{eqnarray}

\begin{figure}[b] %Forced to place figure before conclusions.
    \includegraphics[width=\columnwidth]{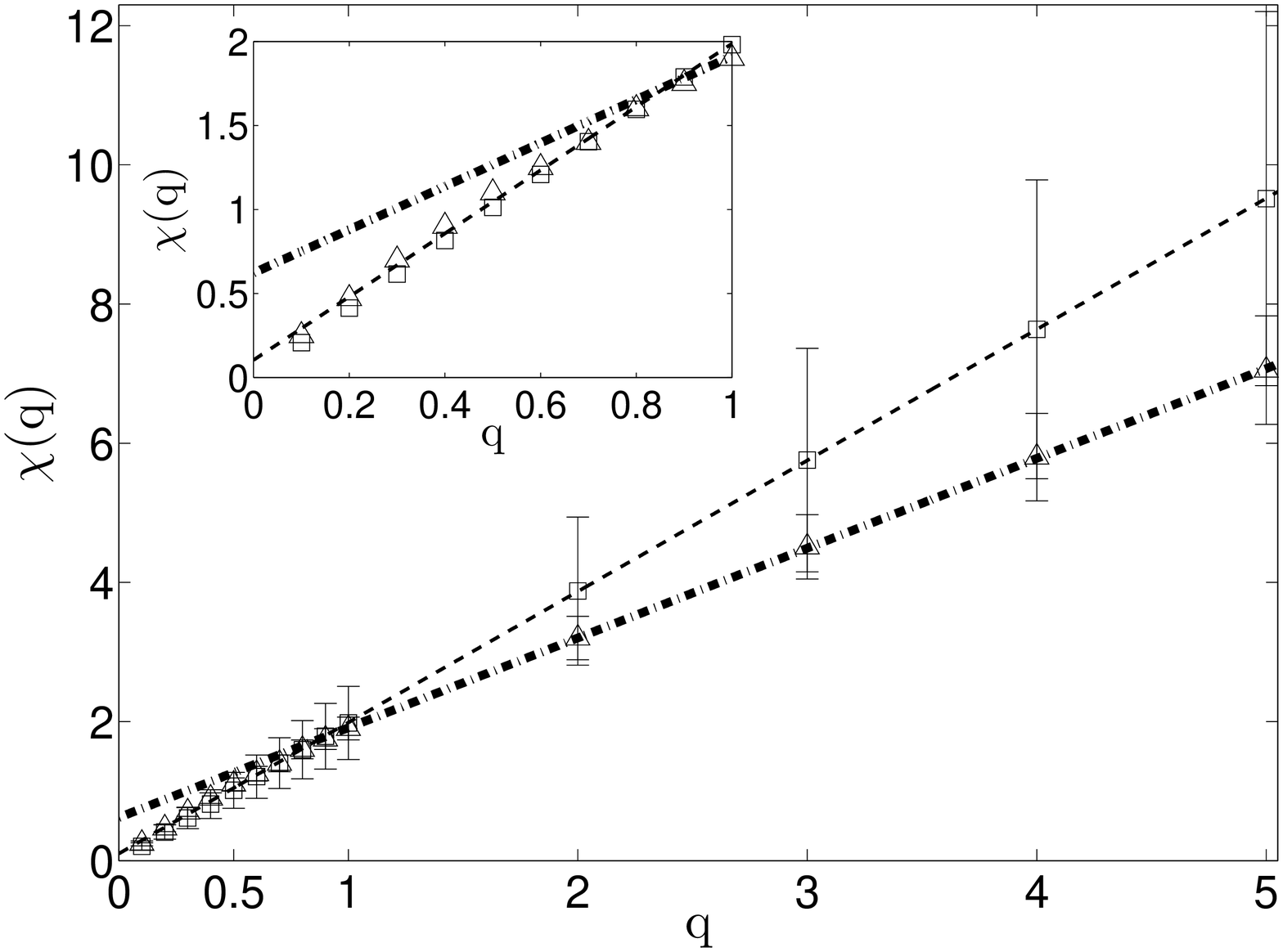}
    \caption{\label{fig:02}Mean Inverse Structure Function for experimental data sets displayed in Fig.\ \ref{fig:01} (convective conditions: $\square$; neutral conditions \cite{Skvortsov_10}: $\triangle$). Error bars correspond to $\pm$ mean standard deviation. The thin dashed (convective) and thick dash-dotted (neutral) lines represent a linear best fit over $1 \le q \le 5$ predicted by (\ref{eqn:E:064}) \cite{Schmitt_05}.}
\end{figure}

Similar to particle dispersion in the neutrally stratified turbulent surface layer, this leads to the conclusion that $\chi (q)$ is a linear function of $q$ \cite{Schmitt_05}. We observe that $\chi (q)$ seems to follow the predicted linear trend for $q \geq 1$, but with a noticeably larger gradient than in the neutral surface layer case ($1.88$ and $1.29$ for the stable and neutral conditions respectively).

As explained in \cite{Celani_04, Boffetta_08}, a non-linear response near the value $q = 1$ (if it exists) can be attributed to a contribution of the slow (differentiable) components of turbulent motion. In our observations we found that this change is usually less profound for the turbulent convective surface layer, see Fig.\ \ref{fig:02}.

\section{Conclusions}

We presented experimental results for passive tracer dispersion in the turbulent convective layer when the dispersion of particles was dominated by buoyancy fluxes. We found that our observations can be intrinsically explained with a two-stage model of tracer dispersion. During the first stage of separation turbulence can be considered as quasi-isotropic and tracer particles obey the standard Richardson model. During the next stage (the long-time limit) particle separation follows the `fast' regime of convective dispersion (predicted by Boligiano-Obukhov scaling). This is markedly different to particle dispersion in the neutrally stratified turbulent surface layer where the standard Richardson regime is followed by the much `slower' (ballistic) rate of separation. These results may have implications for the study of transport phenomenon in various geophysical systems.

\bibliography{particles}

\end{document}